\documentclass[sigconf,authorversion=true]{acmart}

\usepackage{booktabs} 

\usepackage{algorithm}
\usepackage{algpseudocode}
\algdef{SE}[DOWHILE]{Do}{doWhile}{\algorithmicdo}[1]{\algorithmicwhile\ #1}%

\usepackage[colorinlistoftodos]{todonotes}

\setcopyright{rightsretained}

\acmDOI{10.475/123_4}

\acmISBN{123-4567-24-567/08/06}

\acmConference[]{}{}{} 
\acmYear{2017}
\copyrightyear{2017}

\begin{document}
\sloppy

\title{Revisiting Definitional Foundations of Oblivious RAM for Secure Processor Implementations}


\author{Syed Kamran Haider}
\affiliation{\institution{University of Connecticut}}
\email{syed.haider@uconn.edu}

\author{Omer Khan}
\affiliation{\institution{University of Connecticut}}
\email{omer.khan@uconn.edu}

\author{Marten van Dijk}
\affiliation{\institution{University of Connecticut}}
\email{marten.van\_dijk@uconn.edu}


\begin{abstract}
Oblivious RAM (ORAM) is a renowned technique to hide the access 
patterns of an application to an untrusted memory.
According to the standard ORAM definition presented by Goldreich and Ostrovsky, two ORAM access sequences must be computationally indistinguishable if the lengths of these sequences are identically distributed.
An artifact of this definition is that it does not apply to modern ORAM implementations adapted in current secure processors technology because of their arbitrary lengths of memory access sequences depending on programs' behaviors (their termination times).
As a result, the ORAM definition does not directly apply; the theoretical foundations of ORAM do not clearly argue about the timing and termination channels.

This paper conducts a first rigorous study of the standard Goldreich-Ostrovsky ORAM definition in view of modern practical ORAMs (e.g., Path ORAM) and demonstrates the gap between theoretical foundations and real implementations.
A new ORAM formulation which clearly separates out termination channel leakage is proposed.
It is shown how this definition implies the standard ORAM definition (for finite length input access sequences) and better fits the modern practical ORAM implementations.
The proposed definition relaxes the constraints around the stash size and overflow probability for Path ORAM, and essentially transforms its security argument into a performance consideration problem.  To mitigate internal side channel leakages, a generic framework for dynamic resource partitioning has been proposed to achieve a balance between performance and leakage via contention on shared resources.

Finally, a `strong' ORAM formulation which clearly includes obfuscation of termination leakage is shown to imply our new ORAM formulation and applies to ORAM for outsourced disk storage. In this strong formulation constraints are not relaxed and the security argument for Path ORAM remains complex as one needs to prove that the stash overflows with negligible probability.

\end{abstract}

%
%
%
%

\maketitle

\section{Introduction} \label{sec:intro}
Security of private data storage and computation in an untrusted cloud server is a critical problem that has received considerable research attention.
A popular solution to this problem is to use \emph{tamper-resistant hardware} based secure processors including TPM~\cite{arbaugh97secure,VirtualCountersSTC06, tcg-spec04}, 
TPM+TXT~\cite{grawrock-book}, Bastion~\cite{bastion},
eXecute Only Memory (XOM)~\cite{xom-modelcheck,xom-os,xom-2000}, 
Aegis~\cite{aegis_processor, aegis_impl}, 
Ascend~\cite{ascend-stc12}, Phantom~\cite{phantom}, Intel SGX~\cite{intelSGX}, and Sanctum~\cite{sanctum}.
In this setting, a user's encrypted data is sent to the secure processor in the cloud, inside which the data is decrypted and computed upon. 
The final results are encrypted and sent back to the user.
The secure processor chip is assumed to be tamper-resistant, i.e., an adversary is not able to look inside the chip to learn any information.  

While an adversary cannot access the internal state of the secure processor, sensitive information can still be leaked through the processor's interactions with the (untrusted) main memory.
Although all the data stored in the external memory can be encrypted to hide the data values, the memory access pattern (i.e., address sequence) may leak information.
For example, existing work~\cite{Islam12} demonstrates that by observing accesses to an encrypted email repository, an adversary can infer as much as 80\% of the search queries.
Similarly, \cite{ZHUANG04} shows that the control flow of a program can be learned by observing the main memory access patterns which may leak the sensitive private data.

Oblivious RAM (ORAM), first proposed by Goldreich and Ostrovsky~\cite{GO96}, is a cryptographic primitive that completely obfuscates the memory access pattern thereby preventing leakage via memory access patterns.
Significant research effort over the past decade has resulted in more and more efficient ORAM schemes \cite{oramDMP, oram11, GMOT11, GMOT12, oram12a, OsORAM, oram97, SCSL11, SSS12, PathORAM, oram12c}.

Generally speaking, an ORAM interface translates a single logical read/write into accesses to multiple randomized locations.  
As a result, the locations touched in successive logical reads/writes have exactly the same distribution and are indistinguishable to an adversary.
More precisely, according to the original definition of ORAM introduced by Goldreich and Ostrovsky~\cite{GO96}, the ORAM access sequences $\mathsf{ORAM}(A_1)$ and $\mathsf{ORAM}(A_2)$ generated by the ORAM for any two logical access sequences $A_1$ and $A_2$ respectively are computationally indistinguishable if $\mathsf{ORAM}(A_1)$ and $\mathsf{ORAM}(A_2)$ have the same length distribution (where the distribution is over the coin flips used in the ORAM interface).
Almost all follow-up ORAM proposals claim to follow the same definition of ORAM security.

A crucial subtlety regarding the above mentioned ORAM security definition is that it is only applicable to the class of ORAM access sequences whose length is \emph{identically distributed}.
Specifically, two ORAM access sequences $\mathsf{ORAM}(A_1)$ and $\mathsf{ORAM}(A_2)$ may in fact be \emph{distinguishable} if they have different length distributions.
In modern secure processors \cite{ascend-stc12,phantom}, a conventional DRAM controller is replaced with a functionally-equivalent ORAM controller that makes ORAM requests on last-level cache (LLC) misses.
Since a program can have different number of LLC misses for different inputs, the lengths of their corresponding ORAM access sequences is \emph{not} identically distributed, and can leak sensitive information (e.g., locality) via the program's \emph{termination channel} by revealing \emph{when} the program terminates.
Furthermore, the specific ORAM implementations also introduce further variance in the length of ORAM access sequences due to the additional caching/buffering used for performance reasons, e.g., a Path ORAM~\cite{PathORAM} caching the position map blocks for future reuse~\cite{oram-asplos15}.
Hence, the original ORAM definition (\cite{GO96}) does not apply to practical ORAM implementations embraced by the modern secure processors due to their arbitrary distributions of lengths of ORAM access sequences.
In other words, this definition does not clearly separates or includes leakage over the program's termination channel.

Another source of leakage under Goldreich and Ostrovsky's ORAM is the ORAM access timing, i.e., \emph{when} an ORAM access is made.
Since the ORAM requests are issued upon LLC misses, the ORAM access timing strongly correlates with the program's locality and can potentially leak sensitive information via the ORAM \emph{timing channel}.
Periodic ORAM access schemes have been proposed to protect ORAM timing channel~\cite{ascend-stc12,leakage-hpca14}.
Notice, however, that these schemes essentially transform the timing channel leakage into the termination channel leakage.
Completely preventing termination channel leakage without sacrificing performance is a hard problem.
Instead, the leakage can be bounded to only a few number of bits~\cite{leakage-hpca14}.

In this work, we show that   Goldreich and Ostrovsky's ORAM appropriately interpreted for infinite length input access sequences not only implies the standard ORAM definition (\cite{GO96}) for finite length input access patterns, but also separates out termination channel leakage via ORAM access sequences.
The proposed definition bridges the gap between theory and practice in the ORAM paradigm for secure processor technology and also simplifies proving the security of  practical ORAM constructions. 
Specifically, for Path ORAM~\cite{PathORAM}, by leveraging the \emph{background eviction}~\cite{oram-isca13} technique, our definition relaxes the bounds on stash size and stash overflow probability while greatly simplifying the security proof presented in \cite{PathORAM} and yet offering similar security properties.

We also analyze a `strong' ORAM definition stating that two sequences $\mathsf{ORAM}(A_1)$ and $\mathsf{ORAM}(A_2)$ must be computationally indistinguishable if the lengths of the {\em input} sequences $A_1$ and $A_2$ are equal. This definition implicitly includes a form of termination channel obfuscation and is applicable for ORAMs used for remote disk storage. Path ORAM satisfies this stronger definition -- its security proof must now show that the stash over flow probability is negligible (a complex analysis).

The paper makes the following contributions:

\begin{enumerate}
	\item A first rigorous study of the original ORAM definition presented by Goldreich and Ostrovsky, in view of modern practical ORAMs (e.g., Path ORAM), 
	demonstrating the gap between theoretical foundations and real implementations in secure processor architectures.
	\item We show that the Goldreich and Ostrovsky ORAM definition interpreted for infinite length input sequences separates out leakage over the ORAM termination channel leakage. We show
	how  this definition implies the Goldreich and Ostrovsky ORAM definition for finite length input sequences, fits the modern practical ORAM implementations  in secure processor architectures, and greatly simplifies the Path ORAM security analysis by relaxing the constraints around the stash size and overflow probability, 
	and essentially transforms the security argument into a performance consideration problem.
	\item A generic framework for dynamic resource partitioning in secure processor architectures is proposed to control leakage via contention on shared resources, allowing 
	leakage vs. performance trade-offs. In particular, this can be used to reason analyse termination channel leakage.
	\item We analyze a `strong' ORAM definition which implies  the Goldreich and Ostrovsky ORAM definition interpreted for infinite length input sequence. The `strong' ORAM definition implicitly includes obfuscation of the ORAM termination channel and this is useful in ORAM for remote disk storage (in order to prove that Path ORAM satisfies this definition one now needs to show a negligible probability of stash overflow).
\end{enumerate}



\section{Background} \label{sec:background}

\subsection{Leakage Types via Address Bus Snooping} \label{sec:leakages}
Privacy of user's sensitive data stored in the cloud has become a serious concern in computation outsourcing.
Even though all the data stored in the untrusted storage can be encrypted, an adversary snooping the memory address bus in order to monitor the user's interactions with the encrypted storage can potentially learn sensitive information about the user's computation/data~\cite{ZHUANG04,Islam12}.
In particular, such an adversary can potentially learn secret information about the user's program/data by observing the following three behaviors:

\begin{enumerate}
\item The addresses sent to the main memory to read/write data (i.e., \emph{the address channel}).
\item The time \emph{when} each memory access is made (i.e., \emph{the timing channel}).
\item The total runtime of the program (i.e., \emph{the termination channel}).
\end{enumerate}

The countermeasures to prevent leakage via the above mentioned channels are orthogonal to each other and can be implemented as needed.

\subsection{Oblivious RAM} \label{sec:oram}

Oblivious RAM is a renowned technique that obfuscates a user's access pattern to an untrusted storage so that an adversary monitoring the access sequence to the storage cannot learn any information about the user's application or data.
Informally speaking, the ORAM interface translates the user's access sequence of program addresses $A = (a_1,~a_2,~\ldots,~ a_n)$ into a sequence of ORAM accesses $S = (s_1,~s_2,~\ldots,~ s_m)$ such that for any two access sequences $A_1$ and $A_2$, the resulting ORAM access sequences $S_1$ and $S_2$ are computationally indistinguishable given that $S_1$ and $S_2$ are of same length.
In other words, the ORAM physical access pattern ($S$) is independent of the logical access pattern ($A$), except the lengths of the two access patterns which are correlated.
Precisely, an ORAM protects against leakage via the memory address channel only (cf. Section~\ref{sec:leakages}).
The data stored in ORAMs should be encrypted using probabilistic encryption to conceal the data content and also hide which memory location, if any, is updated.
With ORAM, an adversary is not able to tell (a) whether a given ORAM access is a read or write, (b) which logical address in ORAM is accessed, or (c) what data is read from/written to that location.
We revisit the formal definition of ORAM presented by Goldreich and Ovstrofsky~\cite{GO96} and discuss it in more detail in Section~\ref{sec:goldreich}.

\subsection{Path ORAM} \label{sec:basic-pathoram}
Path ORAM~\cite{PathORAM} is currently the most efficient and simplified ORAM scheme for limited client (processor) storage.
Over the past few years, several crucial optimizations to basic Path ORAM have been proposed which have resulted in practical ORAM implementations for secure processor setting.

\begin{figure}[t!]
	\centerline{\includegraphics[width=.9\columnwidth]{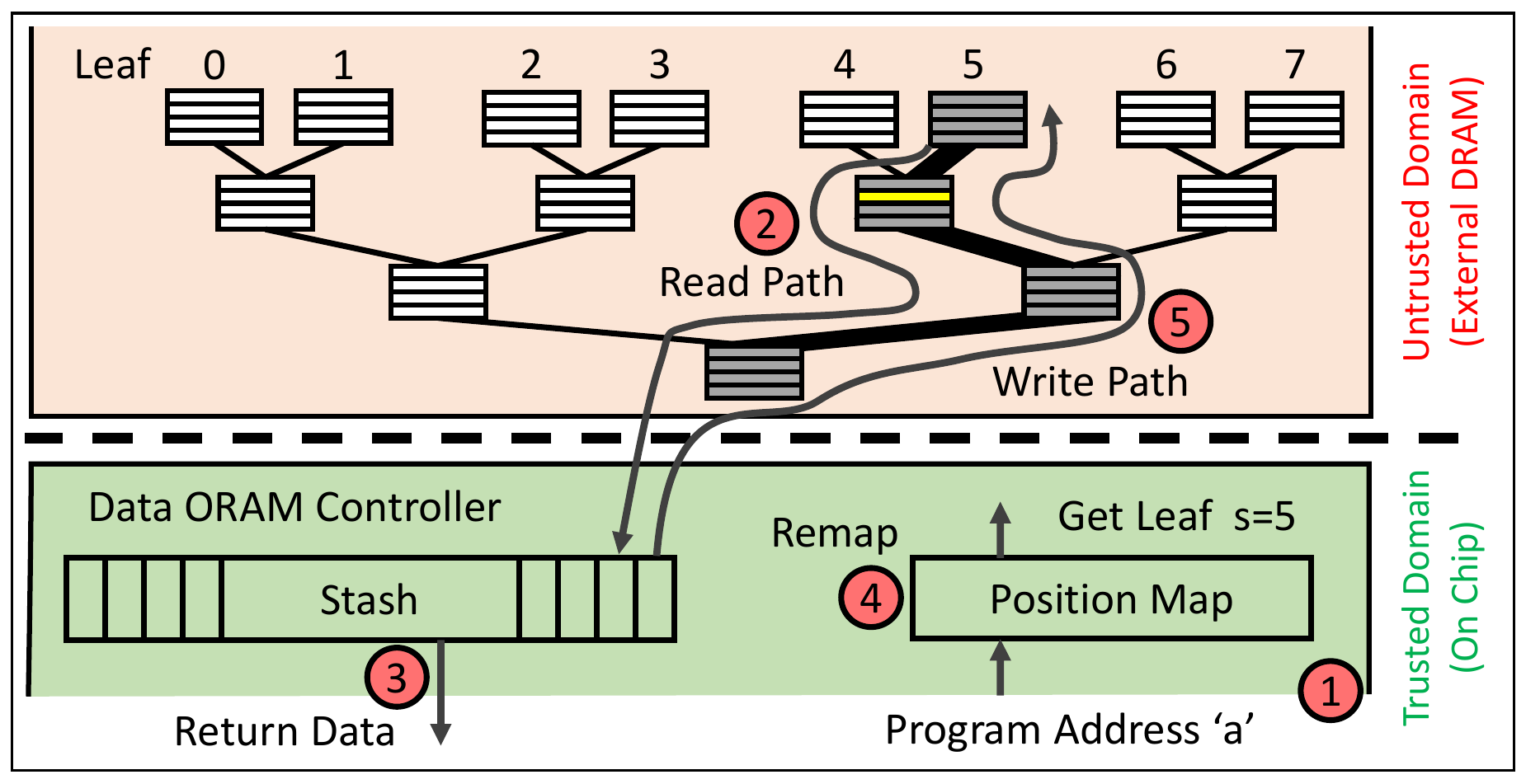}}
	\vspace{-1pt}
	\caption{ A Path ORAM for $L=3$ levels. Path $s=5$ is accessed.  } 
	\label{fig:oram_tree}
	\vspace{-3pt}
\end{figure}

Path ORAM~\cite{PathORAM} has two main hardware components: the 
\emph{binary tree storage} and the \emph{ORAM controller} (cf.  
Figure~\ref{fig:oram_tree}).

\noindent\textbf{Binary tree} stores the data content of the ORAM and 
is implemented on DRAM. Each node in the tree is defined as a 
\emph{bucket} which holds up to $Z$ data blocks.
Buckets with less than $Z$ blocks are filled with \emph{dummy blocks}.
To be secure, all blocks (real or dummy) are encrypted and cannot be 
distinguished.
The root of the tree is referred to as level $0$, and the leafs as level $L$. 
Each leaf node has a unique leaf label $s$.
The path from the root to leaf $s$ is defined as path $s$.
The binary tree can be observed by any adversary and is in this sense 
not trusted.

\noindent\textbf{ORAM controller} is a piece of trusted hardware that 
controls the tree structure.  Besides necessary logic circuits, the 
ORAM controller contains two main structures, a \emph{position map} 
and a \emph{stash}.  The \emph{position map} is a lookup table that 
associates the program address of a data block ($a$) with a path in 
the ORAM tree (path $s$).
The \emph{stash} is a piece of memory that stores up to a small number 
of data blocks at a time.

At any time, each data block in Path ORAM is mapped (randomly) to some 
path $s$ via the position map. Path ORAM maintains the following 
invariant: \emph{if data block $a$ is currently mapped to path $s$, 
then $a$ must be stored either on path $s$, or in the stash} (see 
Figure \ref{fig:oram_tree}). Path ORAM follows the following steps 
when a request on block $a$ is issued by the processor.

\begin{enumerate}
	\item Look up the position map with the block's program address 	$a$, yielding the corresponding leaf label $s$.
	\item Read all the buckets on path $s$.  Decrypt all blocks within the ORAM controller and add them to the stash if they are real (i.e., not dummy) blocks.\label{step:pathread}
	\item Return block $a$ to the secure processor.
	\item Assign a new random leaf $s'$ to $a$ (update the position map).\label{step:remap}
	\item Encrypt and evict as many blocks as possible from the stash to path $s$. Fill any remaining space on path $s$ with encrypted dummy blocks.\label{step:writeback}
\end{enumerate}

Step~\ref{step:remap} is the key to Path ORAM's security. This 
guarantees that a random path will be accessed when block $a$ is 
accessed later and this path is independent of any previously accessed 
random paths (\textit{unlinkability}). As a result, each ORAM access 
is random and unlinkable regardless of the request pattern. 

Although, unlinkability property follows trivially from the construction of Path ORAM, another crucial property to be proven is the negligible stash overflow probability for a small sized stash, i.e., $O(\lambda)$ sized stash for $\lambda$ being the security parameter.

\subsection{Recursive Path ORAM} \label{sec:hier-pathoram}

In practice, the position map is usually too large to be stored in the 
trusted processor. 
Recursive ORAM has been proposed to solve this problem~\cite{SCSL11}. 
In a 2-level recursive Path ORAM, for instance, the original 
position map is stored in a second ORAM, and the second ORAM's 
position map is stored in the trusted processor. The above trick can
be repeated, i.e., adding more levels of ORAMs to further reduce the 
final position map size at the expense of increased latency. The 
recursive ORAM has a similar organization as OS page tables.

\subsection{Background Eviction} \label{sec:background-evict}
In Steps~\ref{step:remap} and \ref{step:writeback} of the basic Path 
ORAM operation, the accessed data block is remapped from the old leaf 
$s$ to a new random leaf $s'$, making it likely to stay in the stash for a while.
In practice, this may cause blocks to accumulate in the stash and finally overflow the stash.  
It has been proven in \cite{PathORAM} that the stash overflow probability is negligible for $Z \geq 6$. 
For smaller $Z$, \emph{background eviction} \cite{oram-isca13} has been proposed to prevent stash overflow.

The ORAM controller stops serving real requests and issues background 
evictions (\textit{dummy accesses}) when the stash is full.
A background eviction reads and writes a random path $s_r$ in the 
binary tree, but does not remap any block.  During the writing back 
phase (Step~\ref{step:writeback} in Section~\ref{sec:basic-pathoram}) 
of Path ORAM access, all blocks that are just read in can at least go 
back to their original places on $s_r$, so the stash occupancy cannot 
increase.  In addition, the blocks that were originally in the stash 
are also likely to be written back to the tree as they may share a 
common bucket with $s_r$ that is not full of blocks.  
Background eviction is proven secure in terms of the unlinkability property in \cite{oram-isca13}. 

\section{Goldreich's Oblivious RAM} \label{sec:goldreich}

Oblivious RAM was first proposed by Goldreich and Ostrofsky~\cite{GO96}.
In this section, we first revisit their definition of ORAM and then discuss its implications on modern real ORAM implementations for secure processor architectures, specifically Path ORAM.

\subsection{Formal Definition} \label{sec:gold_def}
Let $A$ be a sequence of program addresses\footnote{Actually, triples $(o_i,a_i,d_i)$ representing write/read/halt, address, and data send to memory.} $a_1, \cdots, a_i, \cdots$ requested by the CPU during a program execution, and let $\mathsf{ORAM}(A)$ be a probabilistic access sequence to the actual storage such that it yields the \emph{correct} data corresponding to $A$.
Then $\mathsf{ORAM}$ is called an oblivious RAM if it is a probabilitic RAM and satisfies the following definition.

\begin{definition}[Oblivious RAM] \cite{GO96} \label{def:gold_def}
For every two logical access sequences $A_1$ and $A_2$ and their corresponding probabilistic access sequences $\mathsf{ORAM}(A_1)$ and $\mathsf{ORAM}(A_2)$, if $|\mathsf{ORAM}(A_1)|$ and $|\mathsf{ORAM}(A_2)|$ are identically distributed, then so are $\mathsf{ORAM}(A_1)$ and $\mathsf{ORAM}(A_2)$.
\end{definition}

Intuitively, according to Definition~\ref{def:gold_def}, the sequence of memory accesses generated by an oblivious RAM does not reveal any information about the original program access sequence other than its length distribution.
Specifically, this definition only protects against the leakage over memory address channel (cf. Section~\ref{sec:leakages}).

In the above definition we usually interpret $A_1$ and $A_2$ as {\em finite} length sequences implying that $|\mathsf{ORAM}(A_1)|$ and $|\mathsf{ORAM}(A_2)|$ will also be finite length. If infinite length input sequences $A$ are allowed, then the orginal ORAM definition (i.e. Definition \ref{def:gold_def}) turns out to be equivalent to Definition \ref{def:proposed_def} in Section \ref{sec:proposed}. We will argue below why it is important to admit infinite length input sequences.

\subsection{A Bogus ORAM}

Explained below, Definition \ref{def:gold_def} for finite length input sequences invites the construction of a strange `bogus' ORAM in which the access sequence of any probabilistic RAM -- even if it is not oblivious -- can be padded with additional accesses so that it becomes oblivious.  Since the access sequence of a non-oblivious probabilistic RAM is only padded, this reveals information about the input access sequence to the probabilistic RAM. This, of course, breaks our intuitive understanding of what oblivious means. The reason why our construction is oblivious is that the additional padding creates a 1-1 correspondence between the access sequence of the probabilistic RAM and the final length of the access sequence after padding; this allows us to abuse  Definition \ref{def:gold_def} as we essentially code all the information about the access sequence of the probabilistic RAM in the termination channel (the length of the ORAM sequence). This means that each access pattern $A$ will produce a unique length $|\mathsf{ORAM}(A)|$ -- so, there are no two different sequences in Definition \ref{def:gold_def} for our `bogus' construction that will be compared. The bogus construction does not introduce any smartness, it effectively pushes all the work of making the access pattern oblivious to making the termination channel oblivious. This observation will lead to a slightly stronger ORAM definition in Section \ref{sec:proposed} which is independent of the concept of a termination channel, i.e., the length of an ORAM access sequence does not play a role in the new definition (which turns out to be equivalent to  Definition \ref{def:gold_def} for unrestricted and possibly infinite length input sequences).

Algorithm \ref{alg:bog} shows how a (non-oblivious) probablistic RAM ${\mbox{\tt RAM}}^f(.)$ can be padded in order to create an ORAM: Here an input access sequence $A$ to ${\mbox{\tt RAM}}^f(.)$ is finite so that a finite length output sequence ${\mbox{\tt RAM}}^f(A)$ is created which can be {\em uniquely} interpreted as an integer $x$ in line 3.\footnote{A memory access in $A$ is a triple $(op,address, data)$ where $op$ represents a read/fetch/load, write/store, or halt. $op$ can be coded using a non-zero bit sequence of length 2.} The resulting padded ORAM sequence has length $x$, see line 4. This means that if $|\mathsf{ORAMWrapper(RAM}^f)(A_1)|$ and $|\mathsf{ORAMWrapper(RAM}^f)(A_2)|$ are identically distributed, then so are $\mathsf{RAM}^f(A_1)$ and $\mathsf{RAM}^f(A_2)$ (this already shows that only very specific $A_1$ and $A_2$ will result in identically distributed  $|\mathsf{ORAMWrapper(RAM}^f)(A_1)|$ and $|\mathsf{ORAMWrapper(RAM}^f)(A_2)|$). Since line 4 only padds $\mathsf{RAM}^f(A_1)$ and $\mathsf{RAM}^f(A_2)$ with an access sequence taken from some a-priori fixed distribution, the padded access sequences $\mathsf{ORAMWrapper(RAM}^f)(A_1)$ and $\mathsf{ORAMWrapper(RAM}^f)(A_2)$ are identically distributed. We conclude that the bogus ORAM satisfies  Definition \ref{def:gold_def} for finite length input sequences.

The above shows the importance of allowing infinite length input sequences in the ORAM definition. 

\begin{algorithm*}[t!]
	\caption{Bogus ORAM}
	\label{alg:bog}
	\begin{algorithmic}[1]
		\Procedure{ORAMWrapper}{${\mbox{\tt RAM}}^f$}($A$)
		\State Access memory according to ${\mbox{\tt RAM}}^f(A)$
		\State Represent ${\mbox{\tt RAM}}^f(A)$ as a binary bit string and interpret as an integer $x$ which is $\geq$ the number of accesses in ${\mbox{\tt RAM}}^f(A)$
		\State Access memory according to another sequence $A'$ (taken from some a-priori fixed distribution) such that the number of accesses in  ${\mbox{\tt RAM}}^f(A)$ combined with $A'$ is equal to $x$
		\EndProcedure
	\end{algorithmic}
\end{algorithm*}

\subsection{Applicability for Secure Processors} \label{sec:llc-secure-proc}
Modern secure processors~\cite{ascend-stc12,phantom} have embraced Path ORAM interface as a part of their trusted computing base (TCB).
In these implementations, the ORAM controller serves the last level cache (LLC) misses by making ORAM requests to the main memory.
Consider the LLC misses sequence of an execution to be the input ($A$) to the ORAM interface defined in Definition~\ref{def:gold_def}.
In order to conclude  indistinguishability (as per the above definition) of  two ORAM access sequences generated as a result of two different LLC misses sequences (i.e., by running different programs, or running same program with different inputs), the ORAM access sequences must have the same length distribution. 
However, since the LLC misses pattern changes dynamically across various programs and different inputs to the same program~\cite{jaleel2010memory}, it is very unlikely that the corresponding ORAM access sequences of two different executions will have the same length distribution.
In particular, this would leak information about the program behavior through the total runtime of the application (i.e., the termination channel).

Another perspective to look at this fact is that  Definition \ref{def:gold_def} is completely satisfied by only a small class of ORAM access sequences whose lengths are identically distributed.
However, in practice, under the secure processor setting, the lengths of ORAM access sequences can have arbitrary different distributions as discussed earlier.
Furthermore, several optimizations and extensions proposed in the literature for Path ORAM, resulting in better performance/security, introduce further probabilistic variance in the total runtime of the program, i.e., the termination channel.
This, as a result, prevents the ORAM definition under consideration from being directly applicable to secure processors.

\subsection{ORAM Optimizations vs. Program Runtime} \label{sec:optimizations}
In the following discussion, we briefly talk about various optimizations and tricks proposed in the literature that have resulted in more and more efficient and secure Path ORAM implementations.
Each of these techniques typically introduces some amount of variance in the length of the ORAM access sequence as function of the program input, hence, modifying the total runtime of the program that essentially correlates with the given input will leak some information about it.

\subsubsection{Unified Path ORAM \& PLB}
Unified ORAM \cite{oram-asplos15} is an improved and state-of-the-art 
recursion technique to recursively store a large position map. 
It leverages the fact that each block in a position map ORAM stores 
the leaf labels for multiple data blocks that are consecutive in the address space. 
In other words, we can find position maps of several blocks in a single access to the position map ORAM, although only one of them is of interest.
Therefore, Unified ORAM caches position map ORAM blocks in a small cache called \emph{position map lookaside buffer} (PLB) to exploit 
locality (similar to the TLB exploiting locality in page tables).
To hide whether a position map access hits or misses in the cache, 
Unified ORAM stores both data and position map blocks in the same 
binary tree.
Having good locality in position map blocks would result in more PLB hits and overall less number of position map accesses to the Unified ORAM tree, and vice versa.

\subsubsection{ORAM Prefetching}
In order to exploit data locality in programs under Path ORAM, ORAM prefetchers have been proposed~\cite{oram-isca13, yu2015proram}.
At first glance, exploiting data locality and obfuscation seem contradictory:
on one hand, obfuscation requires that all data blocks are mapped to random locations in the memory.
On the other hand, locality requires that certain groups of data blocks can be efficiently accessed together.
However, Path ORAM prefetchers address this problem by (statically/dynamically) creating ``super blocks'' of data blocks exhibiting locality, and mapping the whole super block on the same path.
As a result, a single path read for accessing one particular block yields the corresponding super block which is loaded into the LLC, effectively resulting in a prefetch.
Consequently, good data locality in the program results in more prefetch hits and overall less number of ORAM accesses, and vice versa.

\subsubsection{Timing Channel Protection} \label{sec:oramtiming}
As noted earlier, the ORAM definition does {\em not protect} against leakage over timing channel (cf. Section~\ref{sec:leakages}), i.e., \emph{when} an ORAM access is made.
Periodic ORAM schemes have been proposed to protect the timing channel~\cite{ascend-stc12, leakage-hpca14}.
A periodic ORAM always makes an access at strict periodic intervals, where the time interval $O_{int}$ between two consecutive accesses is public. 
If there is no pending memory request when an ORAM access needs to happen due to 
periodicity, a dummy access will be issued (the same operation as background eviction).
Whereas, if a real request arrives before the next ORAM access time, it waits until the next ORAM access time to enforce a deterministic behavior.
Hence, periodic ORAMs essentially transform the timing channel leakage to the termination channel leakage by potentially introducing extra ORAM accesses due to periodicity.

\subsection{Implications on Path ORAM Stash Size}\label{sec:stash_size}
 Proving that the stash overflow probability is negligible implies Path ORAM's correctness and security.
The stash overflow probability drops exponentially in the stash size.
A significantly complex proof presented in \cite{PathORAM} shows that, for $Z \ge 6$,
a negligible stash overflow probability can be achieved by configuring the stash size appropriately, where $Z$ represents the number of blocks per node in Path ORAM's binary tree.
These parameter settings might be well suited for asymptotic analysis, however, real implementations might choose a different set of parameters to optimize various design points.
For example, a smaller stash size is desired to save hardware area overhead.
Similarly, studies~\cite{oram-isca13} have shown that $Z=3$ yields the best performance for Path ORAM.

For smaller stash sizes and/or $Z<6$, the stash overflow can be prevented through background eviction (cf. Section~\ref{sec:background-evict}) which essentially adds `extra' dummy accesses in the original ORAM access sequence.
Notice, however, that satisfying Definition~\ref{def:gold_def} requires restricting the ORAM access sequences to have identical length distributions, and hence does not apply to background eviction which would probabilistically modify the lengths of ORAM sequences depending upon the stash occupancy which is program input correlated.


As an example, consider a 2-level recursive Path ORAM where the original position map is stored in a second ORAM, and the second ORAM's position map is stored in the trusted processor (cf. Section~\ref{sec:hier-pathoram}). 
Let $A_1$ and $A_2$ be two program address sequences and let $\textsf{ORAM}(A_1)$ and $\textsf{ORAM}(A_2)$ be their corresponding ORAM access sequences.
Notice that each entry of the sequence $\textsf{ORAM}(A_i)$ consists of two accesses corresponding to the position map ORAM and data ORAM respectively, and is therefore likely to increases the stash occupancy by 2.
Further notice that by definition of recursive ORAM structure, each position map ORAM block contains path/leaf labels of several data ORAM blocks consecutively located in the program's address space.

Assume that $A_1$ accesses consecutive data blocks in the program's address space, whereas $A_2$ accesses random data blocks.
Then, subsequent accesses from sequence $\textsf{ORAM}(A_1)$ will exhibit higher temporal locality for position map blocks.
This is because several position map accesses -- corresponding to data blocks consecutive in the program's address space -- will access the same position map block which is likely to be present already in the stash.
Therefore, the stash occupancy will grow at a rate of $<2$ blocks per recursive access.
Whereas, subsequent accesses from $\textsf{ORAM}(A_2)$ exhibit extremely poor temporal locality among position map blocks due to the randomized sequence $A_2$, therefore the stash occupancy will grow at a rate of $\approx 2$ blocks per recursive access.
Consequently, two ORAM accesses sequences 
exhibit two different stash occupancies due to the underlying program's behavior.


\section{Proposed Definition} \label{sec:proposed}

In order to argue about indistinguishability of ORAM access sequences, we interpret Goldreich and Ostrovsky's ORAM definition to also incorporate infinite length input access sequences and this implicitly obfuscates termination channel leakage so that the termination channel cannot be used for leakage in the definition (this separates out the termination channel and invalidates our bogus ORAM as an ORAM).



\begin{definition}[Oblivious RAM for infinite access sequences] \label{def:proposed_def}
For every two logical access sequences $A_1$ and $A_2$ of infinite length, their corresponding (infinite length) probabilistic access sequences $\mathsf{ORAM}(A_1)$ and $\mathsf{ORAM}(A_2)$ are identically distributed in the following sense: For all positive integers $n$, if we truncate $\mathsf{ORAM}(A_1)$ and $\mathsf{ORAM}(A_2)$ to their first $n$ accesses, then the truncations $[\mathsf{ORAM}(A_1)]_n$ and $[\mathsf{ORAM}(A_2)]_n$ are identically distributed.
\end{definition}

Concrete ORAM constructions to-date have the property that  future memory accesses in $A$ do not influence how the oblivious RAM interface accesses memory now: 

 \begin{definition}[Causality] For all $n$, $\mathsf{ORAM}(A)$ extends the access sequence $\mathsf{ORAM}([A]_n)$, where $[A]_n$ is the truncation of $A$ to the first $n$ accesses. 
 \end{definition}

Assuming causality, Definition \ref{def:proposed_def} implies Definition \ref{def:gold_def} for finite length input sequences $A_1$ and $A_2$: Suppose that lengths $|\mathsf{ORAM}(A_1)|$ and $|\mathsf{ORAM}(A_2)|$ are identically distributed. Since $A_1$ and $A_2$ are finite length, also $\mathsf{ORAM}(A_1)$ and $\mathsf{ORAM}(A_2)$ are finite length. Because they are identically distributed there exists a maximum possible length $n$, i.e., $|\mathsf{ORAM}(A_1)|$ and $|\mathsf{ORAM}(A_2)|$ will be $\leq n$.  We may padd $A_1$ and $A_2$ to infinite length sequences $A'_1$ and $A'_2$. Assuming Definition \ref{def:proposed_def} teaches that $\mathsf{ORAM}(A'_1)$ and $\mathsf{ORAM}(A'_2)$ are identically distributed, in particular, $[\mathsf{ORAM}(A'_1)]_n$ and $[\mathsf{ORAM}(A'_2)]_n$ are identically distributed. Due to causality, $[\mathsf{ORAM}(A'_1)]_n$ and $[\mathsf{ORAM}(A'_2)]_n$ extend $\mathsf{ORAM}(A_1)$ and $\mathsf{ORAM}(A_2)$. This implies that $\mathsf{ORAM}(A_1)$ and $\mathsf{ORAM}(A_2)$ must be identically distributed, hence, Definition \ref{def:gold_def} holds for finite length input sequences $A_1$ and $A_2$.

If  in Definition  \ref{def:gold_def} we use  the interpretation of  `identically distributed for infinite length sequences $\mathsf{ORAM}(A_1)$ and $\mathsf{ORAM}(A_2)$' given in Definition \ref{def:proposed_def}, then we may conclude that  Definition \ref{def:proposed_def} is equivalent to  Definition \ref{def:gold_def} for unrestricted and possibly infinite length input sequences $A_1$ and $A_2$.

\subsection{A Stronger Definition}

We can strengthen Definition \ref{def:gold_def} by requiring that $|A_1|$ and $|A_2|$ are identically distributed instead of $|\mathsf{ORAM}(A_1)|$ and $|\mathsf{ORAM}(A_2)|$ being identically distributed:

\begin{definition}[`Strong' Oblivious RAM]  \label{def:strong}
For every two logical access sequences $A_1$ and $A_2$ and their corresponding probabilistic access sequences $\mathsf{ORAM}(A_1)$ and $\mathsf{ORAM}(A_2)$, if $|A_1|$ and $|A_2|$ are 
equal, then so are $\mathsf{ORAM}(A_1)$ and $\mathsf{ORAM}(A_2)$.
\end{definition}

Clearly, this strong definition for unrestricted and possibly infinite length input sequences $A_1$ and $A_2$ implies Definition~\ref{def:proposed_def}  since it covers the case where $|A_1|=|A_2|=\infty$.

Assuming causality, it turns out that the strong Definition \ref{def:strong} restricted to finite length input sequences $A_1$ and $A_2$ also implies Definition~\ref{def:proposed_def}: Suppose that Definition~\ref{def:proposed_def} does not hold and there exist infinite length access sequences $A'_1$ and $A'_2$ and there exists an integer $n$ such that $[\mathsf{ORAM}(A'_1)]_n$ and $[\mathsf{ORAM}(A'_2)]_n$ are not identically distributed. Causality implies that $[\mathsf{ORAM}(A'_1)]_n=[\mathsf{ORAM}([A'_1]_i)]_n$ and $[\mathsf{ORAM}(A'_2)]_n=[\mathsf{ORAM}([A'_2]_j)]_n$ for some (finite) integers $i$ and $j$. Let $k=\max\{i,j\}$. Then (by using causality) $[\mathsf{ORAM}(A'_1)]_n=[\mathsf{ORAM}([A'_1]_k)]_n$ and $[\mathsf{ORAM}(A'_2)]_n=[\mathsf{ORAM}([A'_2]_k)]_n$. We conclude that $\mathsf{ORAM}([A'_1]_k)$ and $\mathsf{ORAM}([A'_2]_k)$ are not identically distributed. This contradicts  Definition \ref{def:strong} for $A_1=[A'_1]_k$ and $A_2=[A'_2]_k$ which both have length $k$.

The next theorem enumerates our findings:

\begin{theorem} \label{theo} Suppose causality. Then, the `strong' ORAM Definition \ref{def:strong} for finite length input sequences implies Goldreich and Ostrovsky's ORAM Definition \ref{def:gold_def} phrased for infinite length input sequences (as in Definition \ref{def:proposed_def}) and this in turn implies Goldreich and Ostrovsky's ORAM Definition \ref{def:gold_def} restricted to finite length input sequences. Our bogus ORAM satisfies Goldreich and Ostrovsky's ORAM Definition \ref{def:gold_def} restricted to finite length input sequences.
\end{theorem}

Finally, we notice that the above definitions can also be adopted in a Universal Composability framework as in \cite{fletcher2016oblivious}.

\subsection{Application}

In the secure processor setting an input sequence $A$ represent the LLC misses sequence. In practice, we may think of the processor to continuously access memory (DRAM) and therefore produce an infinite length input access sequence $A$. This sequence is produced by several programs being contexed switched in and out, some programs terminating and new ones starting. This shows that 
for an ORAM definition to be useful in the secure processor architecture setting we require Goldreich and Ostrovsky's ORAM Definition \ref{def:gold_def} phrased for infinite length input sequences. The termination channel is separated out as the ORAM interface does not terminate and keeps on executing. If a program (module) $P$ terminates it will communicate over a {\em different} I/O channel its computed result.\footnote{Or request input for continuing executing a next program module.} The moment at which this happens leaks information to an observing adversary -- in fact, the adversary can be another program running on the secure processor whose own termination channel leaks into what extent $P$ has slowed down the adversarial program by using shared resources. In Section \ref{ref:term} we propose a framework for analysing leakage over covert channels induced by shared resources. 

In the secure processor architecture setting we do not need the `strong' ORAM definition. It turns out, see Sections \ref{sec:eviction}-\ref{sec:stash}, that PathORAM + background eviction (and other optimizations) satisfies  Goldreich and Ostrovsky's ORAM Definition \ref{def:gold_def} phrased for infinite length input sequences and its security proof is straightforward. However, PathORAM + background eviction does not satisfy the `strong' ORAM definition. We notice that PathORAM without optimizations such as background eviction does satisfy the `strong' ORAM definition and proving this requires a much more complex analysis (as one needs to show that the stash only overflows with negligible probability).  

The `strong' ORAM definition makes sense and is useful in the remote disk storage setting because we will access the remote storage in bursts of requests and we wish the ORAM interface to only reveal the length of the burst and nothing more -- in this way the `strong' ORAM implicitly provides a useful characterization of leakage through the  timing channel (i.e., when accesses happen). 


The above definitions translate to write-only ORAM: HIVE~\cite{hive-ccs14} and Li~{\em et al.} \cite{2013writeonly} essentially use the `strong' write-only ORAM definition since these papers discuss the remote disk storage setting. Flat ORAM~\cite{haider2016flat}, on the other hand, is designed and optimized for the secure processor setting and is secure under the Goldreich and Ostrovsky equivalent of a write-only ORAM definition for infinite length input sequences. Flat ORAM does not  (and does not need to) satisfy the `strong' write-only ORAM.

\subsection{Adapting ORAM Optimizations} \label{sec:eviction}
An ORAM interface satisfying Definition~\ref{def:proposed_def} ``automatically'' caters for the arbitrary and dynamically changing rate of $\mathsf{ORAM}(A)$ accesses to memory per input access in $A$, and is therefore naturally a better fit for practical ORAM implementations (e.g., Path ORAM) in the secure processor setting when compared to Goldreich and Ostrovsky's ORAM Definition \ref{def:gold_def} for finite length input sequences:
The cumulative effect of various performance optimizations outlined in Section~\ref{sec:optimizations} on the termination channel can be incorporated in the proposed ORAM by definition.
E.g., the additional accesses added by the periodic ORAM schemes in order to hide the ORAM access timing, or ORAM prefetching resulting in a reduced number of accesses only results in an altered access sequence, which still remains infinite length.

\subsection{Simplified Stash Analysis} \label{sec:stash}
Another crucial advantage of the proposed definition is that it greatly simplifies the stash size analysis for Path ORAM.
As mentioned earlier, the stash must never overflow for the correctness or security of Path ORAM under the `strong' ORAM Definition \ref{def:strong}, which imposes certain restrictions on the minimum stash size and ORAM parameters, e.g., $Z \ge 6$.
Whereas, according to Definition \ref{def:proposed_def}, it is totally acceptable to have a substantial percentage of background eviction accesses among the overall ORAM accesses, if needed, in order to prevent stash overflow for arbitrary parameter settings.
However, the impact of this relaxed ORAM definition with any chosen parameter settings is reflected in the overall performance of the system.
The system performance can then essentially be benchmarked to tune the optimum settings for desired design points depending upon the application.

\section{Privacy Leakage Analysis} \label{ref:term}
Recall from Section~\ref{sec:gold_def} that a standard oblivious RAM protects only against leakage over the memory address channel.
In this section, we first discuss common mitigation techniques for other leakage sources, e.g., the ORAM timing channel and termination channel.
Later, we present a generic framework, called PRAXEN, that offers security vs. performance trade-offs against a wide range of hardware side channel attacks in a secure processing environment.

\subsection{Timing Channel}
\subsubsection{Static Periodic Behavior}
A straightforward approach to hide the ORAM timing behavior is to use a periodic ORAM scheme~\cite{ascend-stc12}, as introduced in Section~\ref{sec:oramtiming}.
An ORAM access is made strictly after predefined periods, whereas the access period is statically defined offline, i.e., before the program runs.

The security of this approach follows trivially as it completely trades off the timing channel leakage with the total runtime of the program, i.e., altering the termination channel behavior.
Notice that even if the periodic ORAM controller dynamically changes some internal performance parameters, such as prefetching rate and threshold to control background evictions rate, the resultant ORAM access sequence being strictly periodic only alters the termination time of a program.

\subsubsection{Dynamic Periodic Behavior}
While the static periodic approach discussed above is secure, studies have shown that this approach can potentially result in significant performance overheads across a range of programs~\cite{leakage-hpca14}.
On one hand, a \emph{constant} rate of ORAM accesses throughout the program execution is desirable for security, whereas on the other hand, a \emph{dynamically varying} access rate is desirable for performance.
In order to achieve a balance between the two extremes, \cite{leakage-hpca14} proposes a framework that splits the program execution into coarse grained (logical) time \emph{epochs}, and enforces, within each epoch, a strict ORAM access rate that is selected dynamically at the start of each epoch.

Let $L_{max}$ be the maximum program runtime in terms of the number of ORAM accesses such that \emph{all} programs can complete with $\le L_{max}$ ORAM accesses.
Let ${\mathcal E}$ denote the list of epochs of a program execution, or the \emph{epoch schedule}, where each epoch is characterized by its number of ORAM accesses, and let ${\mathcal R}$ denote the list of allowed ORAM access rates.
While running a program during a given epoch, the secure processor is restricted to use a single ORAM access rate, and picks a new rate configuration at the start of the next epoch.
Given $|{\mathcal E}|$ epochs and $|{\mathcal R}|$ rates, there are $|{\mathcal R}|^{|{\mathcal E}|}$ possible epoch schedules -- which can potentially reveal the dynamic behavior of the program.
Thus, the timing channel leakage alone can be upper bounded by $\log_2(|{\mathcal R}|^{|{\mathcal E}|}) = {|{\mathcal E}|} \log_2 {|{\mathcal R}|}$ bits.
To control the amount of leakage, ${|{\mathcal E}|}$ can be set to a small value, e.g., ${|{\mathcal E}|} = \log_2 L_{max}$, resulting in only $\log_2 L_{max} \cdot \log_2 |{\mathcal R}|$ bits leakage while achieving good performance.

%
%
%
%

%
%
%

\subsection{Termination Channel}\label{sec:term_leakage_analysis}
If the results of a program are sent back as soon as the application actually terminates, i.e., the actual termination time is visible to the adversary, sensitive information about the application's input can be leaked by this behavior.
Given the maximum number of ORAM accesses $L_{max}$ within which all programs can terminate, the maximum number of termination traces/lengths that any program can possibly have is upper bounded by $L_{max}$, i.e., one trace per termination point.
Therefore, applying the information theoretic argument from~\cite{smith_theory_information_flow_fossacs09, leakage-hpca14}, at most $\log_2 L_{max}$ bits about the inputs can leak through the termination time alone per execution.
In practice, due to the logarithmic dependence on $L_{max}$, termination time leakage is small.
For example, $\log_2 L_{max}=62$ should work for all programs, which is very small if the user's input is at least a few kilobytes.
Further, we can reduce this leakage through discretization of runtime.
For example, if we ``round up'' the termination time to the next $2^{30}$ accesses, the leakage is reduced to lg $2^{62} -30 = 32$ bits.
The overall leakage by both timing and termination channels can be given by $\log_2 L_{max} \cdot \log_2 |{\mathcal R}| + \log_2 L_{max}$ bits.


\subsection{Other Hardware Side Channels}
While an outside adversary can only monitor an ORAM's external side channels, such as timing/termination channels; in a modern multi-core secure processor, there also exist several internal hardware-based side channels due to the inevitable sharing of various structures.
SVF~\cite{svf_isca12} experimentally measured information leakage in a processor and showed that any ``shared structure'' can leak information. 
In particular, privacy leakage over a shared cache has been explicitly demonstrated in \cite{cross-vm-attacks,ristenpart2012}
for two VMs sharing a cache (without TEE support) showing that secret key bits can leak from one VM to the other, even if the VMs are placed on different cores in the same machine.

Researchers have explored how to counter timing channel attacks due to cache interference~\cite{cache_sc_isca07, cache_sc_nomo} where solutions either rely on static or dynamic cache partitioning. 
The static approach lowers processor efficiency but has a strong security guarantee: no information leakage. 
Current solutions based on the dynamic cache partitioning approach improve processor efficiency but do not guarantee bounds on information leakage. 
We note that efficient cache partitioning is important as it improves processor efficiency~\cite{cache_partitioning_srini_sc04, Qureshi:2006:UCP, pipp_isca09, cloud_cache_hpca11, vantage_micro12, jigsaw_pact13, ubik_asplos14}.

Researchers have also explored how to counter timing channel attacks due to network-on-chip interference in multi-cores~\cite{noc_sc_suh_nocs12, surf_noc_isca13}. 
Both these schemes use static network partitioning to enable information-leak protection through the processor communication patterns. 

Finally, the most important shared resource channel in the ORAM context that leaks information from the hardware layer is the shared ORAM controller that connects (via a traditional memory controller) the processor to the off-chip memory. 
A recent work \cite{timing_attack_host} shows that, under Path ORAM, an adversary running a malicious thread at one of the cores of the multi-core system can learn sensitive information about the behavior of user thread(s) running on other core(s) by introducing contention at the shared ORAM controller and observing the service times of its own requests.
Again, a static partitioning scheme for this information leakage channel can be used at the cost of efficiency.

We want to design a generic dynamic resource partitioning scheme, applicable to any shared resource(s), based on the insight that leakage can be quantified using information theory~\cite{smith_theory_information_flow_fossacs09, theory_predictive_bb_timing_channels_ccs10, theory_predictive_mitigation_timing_channels_ccs11}, in order for achieving a balance between security and performance. 

\subsection{PRAXEN: A PRivacy Aware eXecution ENvironment}

\begin{algorithm*}[t!]
	\caption{Resource Scheduling}
	\label{alg:sch}
	\begin{algorithmic}[1]
		\Procedure{PrivacyAwareScheduler}{${\mathcal F}$} 
		\While {True} 
		\If {$Time \in [NextDecisionPoint, NextDecisionPoint +\delta]$} \Comment{$\delta$ makes the approach reliable}
		\State Obtain $PerfInd_i$ for thread $i$ corresponding to NextDecisionPoint
		\State $x\leftarrow {\mathcal F}(PastHist,NextDecisionPoint,PerfInd_i)$
		\State $PastHist += \{x\}$
		\State Change configuration $c_i$ to the one indicated in $x$ at time $NextDecisionPoint +\delta$
		\State $NextDecisionPoint = Next(PastHist,Time)$
		\EndIf
		\EndWhile
		\EndProcedure
	\end{algorithmic}
\end{algorithm*}

In order to control privacy leakage  while still dynamically sharing resources for efficiency, we propose a generic resource scheduling strategy which only takes a small, yet a sufficient, amount of information about the current and past execution of application threads into account. 

Each application thread $i\in A$  is associated with a configuration $c_i$ which serves as input to the resource scheduler for allocating resources to each thread, i.e., the scheduler assigns resources to each thread according to some (probabilistic) algorithm 
$$Alloc((c_i)_{i\in A}).$$
For example, based on the collection of configurations $(c_j)_{j\in A}$, the resource scheduler may first, by using interpolation and extrapolation, reconstruct a complete approximate picture of all performance indicators which measure how all resources are being used by each of the threads. 
This rough picture is used to allocate the current resources to each thread -- this allocation will not change (it is static) until one of the application thread's configurations $c_i$ changes.  
The reason not to use current measured performance indicators (as an input in $Alloc$) for scheduling is because these dynamically change with respect to execution decisions based on each application thread's state and this gives an uncontrolled amount of leakage. 
As we will see, the above static allocation allows precise control of privacy leakage from one application to another.

We call a change of thread $i$'s configuration from a current configuration $c_i$ to a new configuration $c'_i$ a decision point for $i$. 
Each decision point is associated with an actual time $t_i$. 
At a decision point, the scheduler takes the real, i.e. actual measured, performance indicators of thread $i$ in combination with its history of resource allocations to select a new configuration $c'_i$ together with

\begin{itemize}  
\item A future time $t'_i$ at which the next decision point for $i$ occurs, as well as 
\item A set of future configurations $C'_i$ from which the next configuration for $i$ will be taken.
\end{itemize}

We record the tuples $(i, c_i, t_i, C_i, t'_i)$ in a history ordered by time $t_i$. 
Notice that according to this ordering $(i,c_i,t_i,C_i,t'_i)<(i,c'_i,t'_i,C'_i,t''_i)$ and the above requirements state
\begin{equation} c'_i \in C_i. \label{eq:cond} \end{equation}
So, at time $t_i$ a decision has been made about \emph{what} configuration for $i$ can be selected at the next decision point, and \emph{when} this decision is applied. 

For a current time $t$ we can extract from the past history $PastHist$ of decision points the most recent tuples $(i,c_i, t_i, C_i, t'_i)$ with $t_i<t$ for $i\in A$. 
We compute the time of the next upcoming decision point $NextDecisionPoint$ as
$$ Next(PastHist,t) = \min_{i\in A,  t'_i\geq t} t'_i.$$
Let $i$ be the application thread which corresponds to the upcoming decision point. 
At this decision point the scheduler is allowed to only change $i$'s configuration: 
The scheduler computes
\begin{equation}(i,c'_i,t'_i,C'_i,t''_i) \leftarrow {\mathcal F}(PastHist_i,NextDecisionPoint,PerfInd_i)\label{eqF}\end{equation}
where $PastHist_i$ represents the past history of decision points of thread $i$ and $PerfInd_i$ represent the (history of) {\em measured performance indicators of only thread $i$}. 
Here ${\mathcal F}$ satisfies (\ref{eq:cond}) and $t'_i = NextDecisionPoint$. 
If the scheduler decides not to change $i$'s configuration, then $c'_i=c_i$ in (\ref{eqF}).
Our approach is formalized in Algorithm \ref{alg:sch}.

\vspace{10pt} \noindent
\textbf{Leakage Analysis:} In the worst case all cores/threads, except for one, can collaborate (i.e., act as malicious threads) to observe  one specific (victim) thread $i$ (and, in particular, observe its  configuration changes).
Note that the collaborating threads can \emph{only} observe the victim thread $i$ through changes in resource allocation. We argue that this information is fully captured by $i$'s {\em configuration} changes and the times when these changes happened:
The reason is that each epoch  has  (1) a static resource allocation among threads -- e.g., DRAM bandwidth, ORAM access rate etc. -- preventing internal side channel leakages within an epoch, and (2) \emph{indistinguishability} of real vs. dummy ORAM accesses -- preventing external side channel leakages within an epoch.
Therefore,  {\em accros time}  the collaborating threads can only observe and use the output of $Alloc((c_j)_{j\in A})$ in order to extract information about thread $i$. Hence, the privacy leakage of thread $i$  is at most the information about thread $i$ contained in $PastHist$ which includes the history  of configurations (that form the inputs to $Alloc$).
We notice that each decision point at time $t$ in $PastHist$ is the result of an algorithm ${\mathcal F}$ which $only$ takes as inputs  a history $PastHist_j$ of past decision points before time $t$  together with the corresponding $NextDecisionPoint$ (which is also a function of past decision points before time $t$), and  $PerfInd_j$. Therefore, by using induction on $t$, we can prove that only the decision points corresponding to thread $i$ in $PastHist$ contribute to leakage (through $PerfInd_i$) of thread $i$.
We conclude that  privacy leakage of a specific thread $i$ is at most the information about thread $i$ given by the history of $i$'s decision points (i.e., configuration changes and the times at which these happen): The number of leaked bits is at most Shannon entropy
 $$H(PastHist_i) = H(\{(i,c^{(j)}_i,t^{(j)}_i,C^{(j)}_i,t^{(j+1)}_i)\}\subset PastHist ) $$
 and can be bounded as follows: 

In order to compute (\ref{eqF}) assume that ${\mathcal F}$ 
\begin{itemize}
\item First computes the new configuration $c'_i$ based on inputs $PastHist_i$, $NextDecisionPoint$, and $PerfInd_i$ and
\item Next computes the new sets of possible future configurations $C'_i$ and possible future decision points $t''_i$ based on inputs $PastHist_i$, $NextDecisionPoint$, and $c'_i$ (but not $PerfInd_i$ otherwise an upper bound cannot be proven).
\end{itemize}




We may order the random variables $\{(i,c^{(j)}_i,t^{(j)}_i,C^{(j)}_i,t^{(j+1)}_i)\}$ in $PastHist$ as follows:
\begin{eqnarray*}
&&  H(\{(i,c^{(j)}_i,t^{(j)}_i,C^{(j)}_i,t^{(j+1)}_i)\}\subset PastHist ) \\
&=& H\Big( (C^{(n)}_i,t^{(n+1)}_i), c^{(n)}, (C^{(n-1)}_i,t^{(n)}_i), c^{(n-1)}, \ldots, \\
&&  \ldots, (C^{(0)}_i,t^{(1)}_i), c^{(0)} \Big) \\
&=& \sum_{j=0}^n H\Big((C^{(j)}_i,t^{(j+1)}_i) | c^{(j)}, (C^{(j-1)}_i,t^{(j)}_i), c^{(j-1)},  \ldots, \\ 
&&  \ldots, (C^{(0)}_i,t^{(1)}_i), c^{(0)} \Big) + \\
&&  \sum_{j=0}^n H\Big(c^{(j)} |  (C^{(j-1)}_i,t^{(j)}_i), c^{(j-1)},  \ldots, (C^{(0)}_i,t^{(1)}_i), c^{(0)} \Big) \\
&\leq& 0 + \sum_{j=0}^n \log |C^{(j-1)}|
\end{eqnarray*}
where the first sum equals 0 because ${\mathcal F}$ computes $(C^{(j)}_i,t^{(j+1)}_i)$ as a function of $c^{(j)}$ and $PastHist_i$ (up to moment $t^{(j)}_i$); and the second sum is upper bounded by $\log |C^{(j-1)}|$ because $c^{(j)}\in C^{(j-1)}$.
Let $ \lambda_j = \log |C^{(j-1)}|$ then the $i_{th}$ thread leaks at most $\sum_j \lambda_j$ bits.

Given that algorithms ${\mathcal F}$ and $Alloc$ have enough freedom to reallocate resources, our framework offers a controlled leakage model while maintaining optimum performance.
This methodology can be used on almost all resource sharing paradigms.
It particularly has applications in settings where there is a finite bounded leakage budget. 

%
%
%

\section{Conclusion}
We present a first rigorous study of the original oblivious RAM definition presented by Goldreich and Ostrovsky, in view of modern practical ORAMs (e.g., Path ORAM), and demonstrate the gap between theoretical foundations and real ORAM implementations.
Goldreich and Ostrovsky's  ORAM definition appropriately interpreted for infinite length input access sequebces separates out the ORAM termination channel and fits 
modern practical ORAM implementations in the secure processor setting.
The proposed definition greatly simplifies the Path ORAM security analysis by relaxing the constraints around the stash size and overflow probability, 	and essentially transforms the security argument into a performance consideration problem.
A generic framework for dynamic resource partitioning has also been proposed, which mitigates the sensitive information leakage via internal hardware based side channels -- such as contention on shared resources -- with minimal performance loss.

\begin{acks}
The work is partially supported by NSF grant CNS-1413996 for MACS: A Modular Approach to Cloud Security.
\end{acks}

\bibliographystyle{ACM-Reference-Format}
\bibliography{sections/refs,sections/main,sections/generic,sections/security,sections/onChipNetwork,sections/local,sections/hw,sections/bib-sharon}


\end{document}